\pdfoutput=1

\documentclass[nofootinbib,twocolumn,superscriptaddress,aps,preprintnumbers,amsmath,amssymb,prd]{revtex4-1}

\usepackage{mathrsfs}
\usepackage{natbib}
\usepackage{verbatim}
\usepackage{enumerate}
\usepackage{graphicx,bbm,amsbsy,amsfonts,amssymb,amsmath}
\usepackage{bm}
\usepackage{hyperref}
\usepackage{booktabs}
\usepackage{gensymb}
\usepackage{subfigure}
\usepackage{xcolor}
\usepackage{footmisc}
\usepackage{cancel}
\usepackage{blindtext}
\usepackage{ytableau}

\def\2{{(2)}}
\def\1{{(1)}}
\def\0{{(0)}}
\def\m1{{(-1)}}

\let\vec=\mathbf

\let\oldAA\AA
\renewcommand{\AA}{\text{\normalfont\oldAA}}

\hypersetup{
    colorlinks=true,       % false: boxed links; true: colored links
    linkcolor=red,          % color of internal links (change box color with linkbordercolor)
    citecolor=blue,        % color of links to bibliography
    filecolor=magenta,      % color of file links
    urlcolor=blue           % color of external links
}

\newcommand{\be}{\begin{equation}}
\newcommand{\ee}{\end{equation}}
\newcommand{\ea}{\end{eqnarray}}
\newcommand{\ba}{\begin{eqnarray}}
\begin{document}

%%%%%%%%%%%%%%%%%%%%%%%%%%%%%%%%%%%%%%%%%%%%%%%%%%%%%%%%%%%%%%%%%%%%%%%%%%%%%%%%%%%%%%%%%%%%%%%%%%%%

\preprint{UMN-TH-4536/26}
\preprint{FTPI-MINN-26-16}

\title{Suppressing Extra-Dimensional Axion Isocurvature Dynamically}

\author{Gongjun Choi}
\email{choi0988@umn.edu}
\affiliation{William I. Fine Theoretical Physics Institute, School of Physics and Astronomy,\\
University of Minnesota, Minneapolis, Minnesota 55455, USA}

\author{Tony Gherghetta}
\email{tgher@umn.edu}
\affiliation{School of Physics and Astronomy, University of Minnesota, Minneapolis, Minnesota 55455, USA}

%%%%%%%%%%%%%%%%%%%%%%%%%%%%%%%%%%%%%%%%%%%%%%%%%%%%%%%%%%%%%%%%%%%%%%%%%%%%%%%%%%%%%%%%%%%%%%%%%%%
\begin{abstract}
Extra-dimensional QCD axion is well motivated by string compactifications and enjoys enhanced protection against quality-violating effects. If present during inflation, however, its quantum fluctuations generate isocurvature perturbations that strongly constrain the inflationary scale. We propose a dynamical suppression mechanism in warped five-dimensional models, where a radion-inflaton coupling sets the radion minimum at small inter-brane separation during inflation, temporarily enhancing the effective four-dimensional axion decay constant. After inflation, the radion minimum shifts to larger separation, restoring the standard QCD axion window. In a warped orbifold GUT with Goldberger-Wise stabilization, this mechanism can satisfy CMB isocurvature bounds while allowing substantially higher inflationary scales than in conventional pre-inflationary axion cosmology.
\end{abstract}

%%%%%%%%%%%%%%%%%%%%%%%%%%%%%%%%%%%%%%%%%%%%%%%%%%%%%%%%%%%%%%%%%%%%%%%%%%%%%%%%%%%%%%%%%%%%%%%%%%%%

\date{\today}
\maketitle

%%%%%%%%%%%%%%%%%%%%%%%%%%%%%%%%%%%%%%%%%%%%%%%%%%%%%%%%%%%%%%%%%%%%%%%%%%%%%%%%%%%%%%%%%%%%%%%%%%%%

\section{Introduction}

The QCD axion offers a well-motivated solution to the strong CP problem and can also account for the observed dark matter abundance~\cite{Baker:2006ts,Preskill:1982cy,Abbott:1982af,Dine:1982ah}. In conventional field-theoretic constructions, it arises as the pseudo--Nambu--Goldstone boson of a spontaneously broken global $U(1)_{\rm PQ}$ symmetry anomalous under $SU(3)_c$~\cite{Peccei:1977hh,Peccei:1977ur,Weinberg:1977ma,Wilczek:1977pj}. Nonperturbative QCD effects generate an axion potential whose minimum dynamically relaxes the effective $\bar\theta$ angle to zero, consistent with the CP-conserving vacuum of vectorlike QCD~\cite{Vafa:1984xg}.

This picture is vulnerable to the axion quality problem where additional contributions to the axion potential must be suppressed to extraordinary accuracy in order not to shift the minimum away from the CP-conserving point~\cite{Kamionkowski:1992mf,Holman:1992us,Barr:1992qq,Ghigna:1992iv}. Extra-dimensional axions, which can arise as zero modes of higher-form gauge fields in compactifications~\cite{Witten:1984dg,Choi:1985je,Barr:1985hk,Choi:2003wr,Dvali:2005an,Flacke:2006ad,Arvanitaki:2009fg,Cox:2019rro,Bonnefoy:2020llz,Gherghetta:2020keg,Lee:2021slp,Reece:2023czb,Burgess:2023ifd,Choi:2023gin,Craig:2024dnl,deGiorgi:2024elx,Reece:2025thc,Petrossian-Byrne:2025mto,Benabou:2025kgx,Choi:2025lkg,Loladze:2025uvf,Agrawal:2025rbr,Chakraborty:2025lyp,Csaki:2026qjl,Choi:2026kxu,Hor:2026eyx}, offer a natural way to address this issue. In such theories, higher-form gauge invariance~\cite{Dvali:2005an,Burgess:2023ifd,Choi:2023gin} and higher-form global symmetries~\cite{Reece:2023czb,Craig:2024dnl,Reece:2025thc,Petrossian-Byrne:2025mto,Choi:2025lkg,Loladze:2025uvf,Choi:2026kxu} tightly constrain the allowed interactions, protecting the axion potential without relying on an accidental global 0-form $U(1)_{\rm PQ}$ symmetry.

Extra-dimensional axions, appearing as physical degrees of freedom in the four-dimensional (4D) effective theory without the spontaneous breaking of a global 0-form $U(1)_{\rm PQ}$ symmetry, are naturally present during inflation. As a result, they belong to the pre-inflationary class of axion cosmology, generating isocurvature perturbations that are independent of the adiabatic fluctuations seeded during the inflationary epoch. If the axion is light during inflation, its quantum fluctuations generate angular perturbations
\begin{equation}
\delta \theta_{a,\rm inf} = \frac{\delta a}{f_{\rm inf}}\simeq \frac{H_{\rm inf}}{2\pi f_{\rm inf}}\,,
\end{equation}  
where $f_{\rm inf}$ is the effective axion decay constant during inflation and the axion field fluctuations, $\delta a \simeq H_{\rm inf} / (2\pi)$ with $H_{\rm inf}$ the Hubble expansion rate during inflation. Consequently, the power spectrum of the resulting angular fluctuations takes the form  
\begin{equation}
\mathcal{P}_{\rm iso}(k_{\rm CMB}) = \left(\frac{\Omega_a}{\Omega_{\rm cdm}}\right)^2 \left( \frac{H_{\rm inf}}{\pi f_{\rm inf} \theta_{a,i}} \right)^2 \,,
\label{eq:Piso}
\end{equation}  
where $\Omega_a / \Omega_{\rm cdm}$ represents the fraction of cold dark matter contributed by axions, and $\theta_{a,i}$ is the initial misalignment angle. The amplitude of this primordial isocurvature component is tightly constrained by cosmic microwave background (CMB) observations~\cite{Planck:2018jri},  
\begin{equation}
\beta_{\rm iso} \equiv \frac{\mathcal{P}_{\rm iso}(k_{\rm CMB})}{\mathcal{P}_{\zeta}(k_{\rm CMB}) + \mathcal{P}_{\rm iso}(k_{\rm CMB})} < 0.036 \,,
\label{eq:betaiso}
\end{equation}  
where $\mathcal{P}_{\zeta}(k_{\rm CMB})$ and $\mathcal{P}_{\rm iso}(k_{\rm CMB})$ are the adiabatic curvature and isocurvature power spectra, respectively, evaluated at the CMB pivot scale $k_{\rm CMB}$. 

In the standard inflationary scenario without curvatons, the CMB temperature anisotropies are sourced solely by inflaton fluctuations, yielding a curvature perturbation power spectrum of $\mathcal{P}_{\zeta}(k_{\rm CMB}) \simeq 2.2 \times 10^{-9}$. Under this assumption, the isocurvature bound in (\ref{eq:betaiso}) translates into
\be
\mathcal{P}_{\rm iso}(k_{\rm CMB}) \lesssim 8.2\times 10^{-11}\,.
\label{eq:Pisoconstraint}
\ee
Applying this constraint to (\ref{eq:Piso}) implies that either the axion fraction of cold dark matter must be extremely small, or there must exist a significant hierarchy between the inflationary Hubble scale and the axion decay constant, i.e. $H_{\rm inf}\ll f_{\rm inf}$, assuming a typical initial misalignment angle $\theta_{a,i} \sim 1$. In the latter case, for an axion decay constant near the upper bound of the QCD axion window, $f_{a}\simeq 10^{12}{\rm GeV}$, the isocurvature limit \eqref{eq:betaiso}  requires
\begin{equation}
H_{\rm inf} \lesssim \mathcal{O}(10^7)\,{\rm GeV}\,,
\label{eq:limitHinf}
\end{equation}
assuming no entropy production dilutes the axion abundance before the BBN epoch. Therefore, if the axion decay constant during inflation coincides with its present-day value, the isocurvature constraint \eqref{eq:betaiso} strongly disfavors high inflationary Hubble scales and can force very small slow-roll parameters,
\begin{equation}
\mathcal{P}_{\zeta}(k_{\rm CMB}) \propto \frac{H_{\rm inf}^2}{\epsilon }, \qquad \epsilon \equiv -\frac{\dot{H}_{\rm inf}}{H_{\rm inf}^{2}} \,,
\end{equation}
where $\epsilon$ is the slow roll parameter. This raises an important question: how much can the isocurvature bound (\ref{eq:betaiso}) be relaxed in extra-dimensional axion models, and can inflationary scales far above the conventional limit \eqref{eq:limitHinf} be made viable?

In this work, we explore a resolution of the isocurvature problem for extra-dimensional axions by considering a 1-form axion embedded within a five-dimensional (5D) warped orbifold GUT framework, as proposed in~\cite{Choi:2025lkg} (for the flat extra dimension case, see Ref.~\cite{Benabou:2025kgx}). In this setup, the canonical QCD axion window, $10^{9}{\rm GeV} \lesssim f_a \lesssim 10^{12}{\rm GeV}$, can be made compatible with grand unification of the Standard Model gauge group. This is achieved because the renormalization group evolution of the gauge couplings is dominated by the zero modes of gauge and charged matter fields, which prevents the linear growth of the couplings across the KK mass gap. Consequently, within this framework, one can address the isocurvature problem of the extra-dimensional axion without invoking entropy production to dilute the axion abundance $\Omega_a$.

In this model, the 4D effective axion decay constant takes the form
\ba
f_{a}
=\sqrt{\frac{k}{4g_{5C}^{2}}\,\frac{1}{e^{2\pi kr_{c}}-1}}\simeq \frac{1}{4\sqrt{2}\pi}  \sqrt{N_{\rm CFT}}\,ke^{-\pi kr_{c}},
\label{eq:fa}
\ea
where $g_{5C}$ is the 5D gauge coupling of the bulk $U(1)_C$ gauge group, $k$ is the AdS curvature scale, $r_{c}$ is the compactification radius and $N_{\rm CFT}$ denotes the number of colors in the 4D dual CFT. In obtaining the approximate expression, we used $e^{2\pi kr_{c}}\gg 1$ 
and $8\pi^2/(g_{5C}^2k) \simeq N_{\rm CFT}$.

To realize a temporary large hierarchy between $f_{\rm inf}$ and $H_{\rm inf}$ during inflation in the context of an extra-dimensional axion model based on Randall–Sundrum (RS) warped geometry~\cite{Randall:1999ee,Randall:1999vf}, we consider the possibility that the warp factor during inflation was significantly larger than its post-inflationary value. This scenario can be achieved if the compactification radius during inflation is smaller than its later value after inflation.\footnote{Alternatively, see \cite{Chakraborty:2025lyp,Kawasaki:2015lea,Kitano:2023mra} which suppresses the isocurvature perturbations of an extra-dimensional axion (or axion in an extra-dimensional theory) by temporarily increasing the axion mass above the Hubble scale during inflation.}

For the modulus stabilization, we adopt the Goldberger–Wise mechanism~\cite{Goldberger:1999uk,Goldberger:1999un}, wherein a bulk scalar $\Phi(y)$ with brane-localized vacuum expectation values (VEVs) that have a mild hierarchy generate a radion potential whose minimum determines $r_{c}$. We consider a time-dependent UV boundary condition on $\Phi(y)$, which causes a smaller inter-brane separation during inflation. After inflation, the radion relaxes to its late-time minimum at larger separation. This approach is in the same spirit as dynamical decay constant solutions to the axion isocurvature problem, as discussed in Refs.~\cite{Linde:1991km,Linde:1990yj,Kasuya:1996ns,Kasuya:1997td,Folkerts:2013tua,Kawasaki:2013iha,Higaki:2014ooa,Chun:2014xva,Fairbairn:2014zta,Nakayama:2015pba,Harigaya:2015hha,Kearney:2016vqw,Allali:2022yvx,Conlon:2022pnx}.

The paper is organized as follows.
In Sec.~\ref{sec:largefainf}, we review radion dynamics in a warped inflationary background, including Goldberger--Wise stabilization and de Sitter brane detunings. In Sec.~\ref{sec:smallextraD}, we present the inflation-induced shift of the radion minimum, derive the resulting enhancement of the axion decay constant, and determine the parameter region consistent with radion stabilization and slow roll. In Sec.~\ref{sec:isocurvature}, we show how this enhanced decay constant suppresses CMB-scale axion isocurvature perturbations. In Sec.~\ref{sec:smallscale}, we examine small-scale fluctuations generated during the transition to the late-time decay constant and show that domain-wall formation is avoided for sufficiently high reheating temperature. Finally, we conclude in Sec.~\ref{sec:conclusion}.

%%%%%%%%%%%%%%%%%%%%%%%%%%%%%%%%%

\section{Radion Dynamics and Warped Inflation}
\label{sec:largefainf}

\subsection{Modulus Stabilization}
\label{sec:GW}

In the 5D RS1 framework, the 5D gravitational action with a negative cosmological constant $\Lambda_{5}<0$ is given by
\ba
S_{\rm 5D}&\supset&\int d^{4}x\int_{-\pi}^{\pi}d\phi\sqrt{g}\left(-\frac{1}{2}M_{5}^{3}R-\Lambda_{5}\right)\cr\cr
&-&\sum_{i={\rm UV,IR}}\int d^{4}x\int_{-\pi}^{\pi}d\phi\sqrt{-g_{i}}\,T_{i,c}\,\delta(\phi-\phi_{i})\,,
\label{eq5DSEHaction}
\ea
where $g$ is the determinant of the 5D metric, $g_{i}$ is the determinant of the induced metric on the $i$--brane, $M_{5}$ denotes the 5D Planck mass, $\phi_{\rm UV}=0$ and $\phi_
{\rm IR}=\pi$, $T_{{\rm UV},c}$ and $T_{{\rm IR},c}$ are the (critical) brane tensions of the UV and IR branes, respectively. For $\Lambda_{5}=-6 M_{5}^{3}k^{2}$ and tuned brane tensions satisfying $\Lambda_{5}=-k T_{{\rm UV},c}=k T_{{\rm IR},c}$, the equations of motion derived from (\ref{eq5DSEHaction}) admit the static metric solution~\cite{Randall:1999ee,Randall:1999vf}
\ba
ds^{2}=e^{-2k r_{c}|\phi|}\eta_{\mu\nu}dx^{\mu}dx^{\nu}-r_{c}^{2}d\phi^{2}
= e^{-2ky} dx^{2} - dy^{2}\,,
\label{eq:metric}
\ea
where $r_{c}$ is the radius of the compact extra dimension and the orbifold $S^{1}/Z_{2}$ coordinate $y$ is related to the angular variable $\phi\in[0,\pi)$ through $y\equiv\phi r_{c}$ with $y_{\rm UV}=0$ and $y_{\rm IR}=\pi r_{c}$.

The modulus (or radion) can be stabilized by the Goldberger-Wise mechanism~\cite{Goldberger:1999uk}, which introduces a bulk scalar field $\Phi$ with 5D action
\ba
S_{\rm GW}&=&\int d^{4}xdy\sqrt{g}\left[\frac{1}{2}g^{MN}\partial_{M}\Phi\partial_{N}\Phi-\frac{1}{2}m_{\Phi}^{2}\Phi^{2}\right]\nonumber\\
&-&\sum_{i={\rm UV,IR}}\int d^{4}x\int_{-\pi}^{\pi}d\phi\sqrt{-g_{i}}\,\lambda_{i}(\Phi^{2}-v_{i}^{2})^{2}\delta(\phi-\phi_{i})\nonumber\\
&\equiv &S_{\Phi, 5}+S_{\Phi,{\rm UV}}+S_{\Phi,{\rm IR}}\,,
\label{eq:SGW}
\ea
where $m_{\Phi}$ is the bulk mass of $\Phi$, and the mass dimensions of parameters and fields are $[\lambda_{\rm UV,IR}]=-2$, $[\Phi]=\frac{3}{2}$, $[v_{\rm UV,IR}]=\frac{3}{2}$. The parameters $v_{\rm UV}$ and $v_{\rm IR}$ are the brane-localized VEVs, which in the large $\lambda_{i}$ limit, fix the boundary values
\be
\Phi(y_{\rm IR})=v_{\rm IR}\equiv c_{\rm IR}k^{\frac{3}{2}},\quad\Phi(y_{\rm UV})=v_{\rm UV}\equiv c_{\rm UV}k^{\frac{3}{2}}\,,
\label{eq:BCs}
\ee
where $c_{\rm IR,UV}$ are dimensionless constants. Throughout this work, we assume $v_{\rm UV,IR}^{2}\ll M_{5}^{3}$ to ensure that the backreaction of $\Phi$ on the 5D geometry is negligible.

For $|\alpha|\ll1$, where $\alpha\equiv m_{\Phi}^{2}/(4k^2)$, the classical bulk profile is $\Phi(y)=A_{+} e^{c_{+}y}+A_{-} e^{c_{-}y}$ with $c_{\pm}=2k\pm\sqrt{4k^{2}+m_{\Phi}^{2}}$ and the constants $A_\pm$ are fixed by the boundary conditions (\ref{eq:BCs}). Substituting this solution into the action and integrating over the extra dimension gives the 4D effective potential for $r_{c}$~\cite{Goldberger:1999uk} 
\ba
V_{\rm GW}(r_{c})&=&\alpha c_{\rm UV}^{2}k^{4} -\alpha c_{\rm IR}^{2}k^{4}e^{-4k\pi r_{c}}\\
&+&2(2+\alpha)k^{4}e^{-4k\pi r_{c}}(c_{\rm IR}-c_{\rm UV}e^{-\alpha \pi kr_{c}})^{2}\,.\nonumber
\label{eq:V(d)}
\ea

The fluctuations of the background metric give rise to massless modes in the 4D low-energy effective theory (EFT): the 4D graviton and a scalar mode associated with fluctuations of the inter-brane separation, $\delta r$. It is convenient to encode the background inter-brane separation $r_{c}$ and its spacetime-dependent fluctuation $\delta r(x)$ in terms of a canonically normalized radion field $\sigma(x)$~\cite{Goldberger:1999un}
\be
\sigma(x)\equiv\sqrt{\frac{6M_{5}^{3}}{k}}e^{-\pi kr(x)}\equiv Fe^{-\pi kr(x)}\,,
\ee
where $r(x)=r_{c}+\delta r(x)$. The radion VEV (equivalently, $r_{c}$) is determined by its effective potential
\ba
V_{\rm GW}(\sigma)&=&\alpha c_{\rm UV}^{2}k^{4}-\alpha c_{\rm IR}^{2}k^{4}\left(\frac{\sigma}{F}\right)^{4}\cr\cr
&+&2(2+\alpha)k^{4}\left(\frac{\sigma}{F}\right)^{4}\left[c_{\rm IR}-c_{\rm UV}\left(\frac{\sigma}{F}\right)^{\alpha}\right]^{2}.\,\nonumber\\ 
\label{eq:Vsigma}
\ea

For $\alpha>0$ and $v_{\rm UV}>v_{\rm IR}$, the minimum of (\ref{eq:Vsigma}) up to corrections of order $\sqrt\alpha$ is
\be
\langle\sigma(x)\rangle\equiv\sigma_{\rm min}\simeq F\left(\frac{v_{\rm IR}}{v_{\rm UV}}\right)^{\frac{1}{\alpha}}\,,
\label{eq:sigmaVEV}
\ee
which corresponds to stabilization of the compact extra dimension radius at
\be
\pi r_{c}=-\frac{1}{k}\ln\left(\frac{\sigma_{\rm min}}{F}\right)\simeq\frac{1}{k\alpha }\ln\left(\frac{v_{\rm UV}}{v_{\rm IR}}\right)\,.
\label{eq:kd}
\ee
Thus, a modest hierarchy between $v_{\rm IR}$ and $v_{\rm UV}$ in AdS$_{5}$ is exponentially amplified into a large hierarchy between the AdS curvature $k$ and the characteristic Kaluza–Klein mass scale,
\be
m_{\rm KK}=\pi ke^{-k \pi r_{c}}\simeq\pi k\left(\frac{v_{\rm IR}}{v_{\rm UV}}\right)^{\frac{1}{\alpha}}\,.
\label{eq:KKmassgap}
\ee

On the other hand, for $\alpha<0$, $V_{\rm GW}(\sigma)$ fails to develop a nontrivial global minimum due to the second term in (\ref{eq:Vsigma}) when the brane tensions are tuned to their critical values. However, $V_{\rm GW}(\sigma)$ can develop a nonzero global minimum when $v_{\rm IR}>v_{\rm UV}$, provided the IR brane tension is slightly detuned by $\delta T_{\rm IR}<0$. The existence of a nontrivial minimum requires $|\delta T_{\rm IR}|\geq|\alpha|c_{\rm IR}^{2}k^{4}$, while boundedness from below requires $|\delta T_{\rm IR}|<(4-|\alpha|)c_{\rm IR}^{2}k^{4}$~\cite{Arkani-Hamed:2000ijo,Rattazzi:2000hs,Creminelli:2001th}.

%%%%%%%%%%%%%%%%%%%%%%%%%%%%%%%%%%%%%%%%%%%%%

\subsection{Inflation in a Slice of AdS$_{5}$}
\label{sec:Inflation}

When the UV and IR brane tensions are detuned from their critical RS values—defined as $\delta T_{\rm UV}^{\rm dS}>0$ and $\delta T_{\rm IR}^{\rm dS}<0$ (see (\ref{eq:detuningsforinflation}))-- the static RS1 background can be replaced by a warped de Sitter geometry~\cite{Kaloper:1999sm,Nihei:1999mt,Kim:1999ja,DeWolfe:1999cp}\footnote{See also \cite{Csaki:1999jh,Cline:1999ts,Kanti:1999sz,Cline:1999tq,Mishra:2025ofh} for other time dependent solutions for the bulk geometry.}
\ba
ds^{2}=\widetilde{a}^{2}(y)(dt^{2}-\widetilde{b}^{2}(t)d\vec{x}^{2})- dy^{2}\,.
\label{eq:warpeddS}
\ea
The $55$-component of the 5D Einstein equation fixes the warp factor to be
\be
\widetilde{a}(y)=\frac{\widetilde{H}}{k}\sinh[k(y_{c}-|y|)]\,,
\ee
where $\sinh(ky_{c})=k/\widetilde{H}$ since $\widetilde{a}(y)=1$ for $y=y_{\rm UV}$. The $00$ and $ii$ components then yield a constant expansion rate for the 4D slices:
\be
\frac{\dot{\widetilde{b}}(t)}{\widetilde{b}(t)}=\widetilde{H}=\text{const}\quad\Rightarrow\quad \widetilde{b}(t)\propto e^{\widetilde{H}t}\,,
\ee
where $\widetilde{H}$ is an integration constant. The consistency of this geometry requires that the Israel junction conditions \cite{Israel:1966rt} be satisfied at both the UV and IR boundaries:
\ba
T_{{\rm UV},c}+T_{{\rm UV}}^{(0)}+\delta {T}_{\rm UV}^{\rm dS}&=&6M_{5}^{3}k\coth(ky_{c})+T_{{\rm UV}}^{(0)}\,,\cr\cr
T_{{\rm IR},c}+\delta {T}_{\rm IR}^{\rm dS}&=& -6M_{5}^{3} k\coth[k(y_{c}-y_{\rm IR})]~~
\label{eq:IsraelJunction}
\ea
where $T_{{\rm UV},c}=-T_{{\rm IR},c}=6M_{5}^{3}k$ are the critically tuned tensions of the static RS solution (\ref{eq:metric}) in the absence of detunings and $T_{{\rm UV}}^{(0)}$ represents additional tuning needed to obtain the observed 4D cosmological constant. For large hyperbolic arguments, we can use the approximation $\coth x\approx1+2e^{-2x}$ in (\ref{eq:IsraelJunction}), which allows the detunings to be expressed as
\ba
\delta {T}_{\rm UV}^{\rm dS}&=&12M_{5}^{3}ke^{-2ky_{c}}\simeq3M_{4}^{2}\widetilde{H}^{2}\cr\cr
\delta T_{\rm IR}^{\rm dS}&=&-12M_{5}^{3}ke^{-2k(y_{c}-y_{\rm IR})}\simeq-3M_{4}^{2}\widetilde{H}^{2}e^{2ky_{\rm IR}},\,\nonumber\\
\label{eq:detuningsforinflation}
\ea
where $\sinh(ky_{c})=k/\widetilde{H}\approx e^{ky_{c}}/2$ and $M_{4}^{2}\simeq M^{3}_{5}/k$ were used for the second equalities with $M_{4}=2.4\times10^{18}\,{\rm GeV}$.

To obtain the Hubble expansion rate perceived by an observer residing in a 4D slice at a fixed $y$, we map the metric (\ref{eq:warpeddS}) onto the metric of standard 4D de Sitter space. This transformation gives a $y$-dependent Hubble parameter
\be
H(y)=\frac{k}{\sinh[k(y_{c}-|y|)]}\,.
\label{eq:H4D}
\ee
In particular, $H(y)$ on the UV brane becomes
\be
H(y_{\rm UV})=\frac{k}{\sinh(ky_{c})}=\widetilde{H}\,.
\ee

In summary, a de-Sitter slicing of AdS$_{5}$ is obtained when the UV and IR brane tensions are detuned from their critical RS values by the specific detunings in (\ref{eq:detuningsforinflation}). These detunings determine the inflationary Hubble scale seen on the UV brane. The additional contribution $T_{\rm UV}^{(0)}$ in (\ref{eq:IsraelJunction}) is assumed to satisfy $T_{\rm UV}^{(0)}\ll \delta T_{\rm UV}^{\rm dS}$, so its effect on the inflationary background is negligible and the geometry remains well approximated by (\ref{eq:warpeddS}).

%%%%%%%%%%%%%%%%%%%%%%%%%%%%%%%%%%%%%%%%%%%%%

\section{Inflation-Enhanced Axion Decay Constant}
\label{sec:smallextraD}

\subsection{Time-Dependent Radion Minimum}

During inflation, we take the inflaton ($\chi$) sector to be localized on the
UV brane. The associated inflaton potential $V_{\rm inf}(\chi)$ therefore
contributes directly to the UV-brane tension detuning appearing in
(\ref{eq:detuningsforinflation}). In general, we also allow an additional
UV-brane-localized contribution $U_{\rm UV}(\chi)$, distinct from the slow-roll
potential, whose role is to help satisfy the detuning conditions. We assume
that the symmetries controlling the inflaton sector allow $U_{\rm UV}$ to
depend on $\chi$ only through $V_{\rm inf}(\chi)$, so that during inflation
both $V_{\rm inf}$ and $U_{\rm UV}$ are effectively constant. In applying the
de Sitter junction conditions, these quantities should be understood as their
quasi-de Sitter background values during the slow-roll epoch, for example at
CMB horizon exit. The residual
$\chi$-dependence is retained when estimating slow-roll corrections. The total
UV-brane detuning is then
\be
\delta T_{\rm UV}^{\rm dS}=V_{\rm inf}(\chi)+U_{\rm UV}(\chi)\,.
\label{eq:UVdetuning}
\ee
To realize an inflationary background on 4D slices following
Sec.~\ref{sec:Inflation}, the positive UV-brane detuning
\eqref{eq:UVdetuning} must be accompanied by a negative detuning
$\delta T_{\rm IR}^{\rm dS}$ of the IR-brane tension.
Using (\ref{eq:detuningsforinflation}), evaluated on the quasi-de Sitter
background described above, this is given by
\be
\delta T_{\rm IR}^{\rm dS}
=-\delta T_{\rm UV}^{\rm dS}e^{2ky_{\rm IR}}
=-(V_{\rm inf}(\chi)+U_{\rm UV}(\chi))e^{2kd_{\rm inf}}\,,
\label{eq:irdetuning}
\ee
where $d_{\rm inf}\equiv y_{\rm IR}-y_{\rm UV}$ is the inter-brane separation during inflation. After inflation, the inflation-induced UV contribution \eqref{eq:UVdetuning} relaxes to zero, while the IR-brane contribution \eqref{eq:irdetuning} remains as a constant term in the radion potential. We assume that the late-time brane tensions are adjusted by the usual cosmological constant tuning, so that the post-inflationary vacuum has negligible vacuum energy. The inflationary detunings in (\ref{eq:UVdetuning}) and (\ref{eq:irdetuning}) contribute to the 4D effective scalar potential as
\be
V_{\rm 4D}(\chi,\sigma)\supset \delta T_{\rm UV}^{\rm dS}+\delta T_{\rm IR}^{\rm dS}\left(\frac{\sigma}{F}\right)^{4}\,.
\label{eq:V4Dscalarpart}
\ee

We next also introduce a Goldberger-Wise scalar field $\Phi$. Irrespective of the specific symmetries imposed on the theory—since these must be respected by $V_{\rm inf}(\chi)$—the inflaton can consistently couple to the Goldberger–Wise scalar $\Phi$ through a UV-brane interaction. Therefore, we assume the following 5D action for $\Phi$ and $\chi$ on the UV brane
\ba 
&&S_{\Phi\chi,{\rm UV}}=\int d^{4}x\,d\phi\sqrt{-g_{\rm UV}}\Bigl[\frac{1}{2}g_{\rm UV}^{MN}\partial_{M}\chi\partial_{N}\chi-V_{\rm inf}(\chi)\cr\cr &&-\lambda_{\rm UV}\left(\Phi^{2}-c_{V}^2 V_{\rm inf}(\chi)^{\frac{3}{4}}-v_{\rm UV}^{2}\right)^{2}\Bigr]\delta(\phi-\phi_{\rm UV}),\,
\label{eq:SGW2} 
\ea
where $c_{V}$ is a dimensionless constant. The mass dimensions of parameters in the inflaton sector are $[\chi]=1$ and $[V_{\rm inf}(\chi)]=4$.

\begin{figure}[t]
\centering
\hspace*{-5mm}
\includegraphics[width=0.4\textwidth]{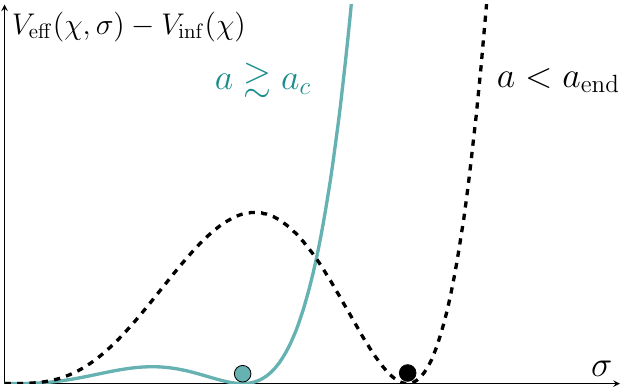}
\caption{Schematic plot showing the $\sigma$-dependent part (radion potential) of $V_{{\rm eff}}(\chi,\sigma)$ in (\ref{eq:Vsigma2}) during inflation $a < a_{\rm end}$ ({\bf{\color{black}{black dashed}}}), and after reheating $a > a_{c}$ ({\bf{\color{teal}{green solid}}}). The radion $\sigma$ tracks the corresponding local minimum given in (\ref{eq:sigmamincases}) for each respective era. 
The post-inflationary reduction of $\sigma_{\rm min}$ increases the inter-brane separation and lowers the axion decay constant $f_a$.
}
\vspace*{-1.5mm}
\label{fig:radpotential}
\end{figure}

The purpose of introducing the modified UV-brane scalar potential is to generate an inflation-induced shift in the UV boundary value of $\Phi$, in addition to $v_{\rm UV}$. From dimensional analysis, a general class of admissible UV-brane operators, assuming a $\Phi \rightarrow -\Phi$ symmetry, can be written schematically as
\be
\sum_{p,q\in\mathbb{Z}}-\lambda_{2}^{q}V_{\rm inf}(\chi)^{\frac{2q+1}{4}}\Phi^{2}+\lambda_{4}^{p}V_{\rm inf}(\chi)^{\frac{p-1}{2}}\Phi^{4}\,,
\label{eq:genericpotential}
\ee
which preserves the symmetries of the inflaton sector with $[\lambda_{2,4}]=-2$.
If a single quadratic and quartic term dominate, the induced minimum scales as
\be
\langle\Phi\rangle\sim\lambda_{2}^{\frac{q}{2}}\lambda_{4}^{-\frac{p}{2}}V_{\rm inf}(\chi)^{(2q-2p+3)/8}\,.
\label{eq:Phivevgeneral}
\ee
The specific choice $(q,p)=(1,1)$ with $\lambda_{2}/\lambda_{4}=c_{V}^2$, corresponding to (\ref{eq:SGW2}), is therefore not unique. Indeed, for $(q,p)\neq(1,1)$, one recovers identical dynamical behavior provided the VEV in (\ref{eq:Phivevgeneral}) matches $c_{V}V_{\rm inf}(\chi)^{3/8}$. So, without loss of generality, we restrict our subsequent analysis to the representative case (\ref{eq:SGW2}) for the sake of simplicity. Accordingly, the 5D action $S_{{\rm GW},\chi}$ for the Goldberger-Wise bulk scalar $\Phi$ and the inflaton field $\chi$ is obtained by modifying the UV-brane part of (\ref{eq:SGW})
\be
S_{{\rm GW},\chi}=S_{\Phi, 5}+S_{\Phi\chi,{\rm UV}}+S_{\Phi,{\rm IR}}\,. 
\label{eq:SGWchi}
\ee

For the convenience of comparing parameters, we express the inflaton potential and the 4D Hubble expansion rate $H_{\rm inf}$ during inflation in units of the AdS curvature $k$, namely\footnote{Note that $H_{\rm inf}$ is determined by the net 4D de Sitter vacuum energy. As the setup contains the Goldberger–Wise scalar $\Phi$ in addition to $V_{\rm inf}(\chi)$, the boundary contribution arising from the bulk action of $\Phi$ partially cancels the UV-brane detuning $\delta T_{\rm UV}^{\rm dS}$ in (\ref{eq:UVdetuning}), leaving $V_{\rm inf}(\chi)$ as the dominant residual contribution. In this sense, the effective UV-brane detuning relevant for inflation is controlled by $V_{\rm inf}(\chi)$. An analogous partial cancellation likewise occurs for the IR-brane detuning $\delta T_{\rm IR}^{\rm dS}$.}
\be
V_{\rm inf}(\chi)=c_{\rm inf}^{8/3} k^{4},
\quad H_{\rm inf}=\sqrt{\frac{V_{\rm inf}(\chi)}{3M^{2}_{4}}}=\frac{c_{\rm inf}^{4/3}k}{\sqrt{3}b_{\rm CFT}}\,,
\label{eq:parametersink}
\ee
where $V_{\rm inf}(\chi)$ is 4D de Sitter vacuum energy, $M_{4}\sim b_{\rm CFT}k$~\footnote{
This scaling follows from the relation
$M_4^2\simeq M_5^3/k$ and the AdS/CFT estimate
$M_5^3/k^3\sim N_{\rm CFT}^2$, which give $M_4/k\sim N_{\rm CFT}\sim b_{\rm CFT}$, up to order-one factors.
} and $b_{\rm CFT}\simeq N_{\rm CFT}$ is the $\beta$-function coefficient of the strongly-coupled 4D dual CFT.

With the choice of $\alpha<0$, a particularly intriguing possibility opens up provided
\be
c_{\rm UV}\ll c_{V} c_{\rm inf}\lesssim c_{\rm IR}\,.
\label{eq:vevhierarchy}
\ee
Suppose $d_{\rm inf}$ remains fixed during inflation with $V''_{\rm GW}(\sigma_{\rm min})\gtrsim H_{\rm inf}^{2}$ satisfied.
Since $H_{\rm inf}$ (and therefore $V_{\rm inf}(\chi)$) is approximately constant over the inflationary epoch, the quantity $c_{V} V_{\rm inf}(\chi)^{\frac{3}{8}}$ in \eqref{eq:SGW2} effectively supplants $v_{\rm UV}$ as the UV boundary value of $\Phi$ during this epoch,\footnote{Since the beginning of inflation is model dependent, a sizable $\chi$ dependence of $\Phi(y_{\rm UV})$ would make $f_{\rm inf}$ at CMB horizon exit ambiguous. We therefore take $\Phi(y_{\rm UV}) \simeq \text{constant}$ during inflation, which makes $f_{\rm inf}$ well defined and maximizes the suppression of $\delta\theta_a$ for fixed parameters. }
i.e. during inflation 
\be
\Phi(y_{\rm UV})=c_{V} V_{\rm inf}(\chi)^{\frac{3}{8}}=c_{V}(3M_{4}^{2}H_{\rm inf}^{2})^{\frac{3}{8}}\,.
\label{eq:enhancedBV}
\ee
After inflation ends, $\Phi(y_{\rm UV})\rightarrow v_{\rm UV}$ with $V_{\rm inf}(\chi)\rightarrow0$.

Assuming the condition (\ref{eq:vevhierarchy}) is satisfied, the 5D setup arising from \eqref{eq:SGW2} yields the following 4D effective scalar potential during the inflationary epoch
\ba
V_{\rm 4D}(\chi,\sigma)&=&
V_{\rm inf}(\chi)+U_{\rm UV}(\chi)+\delta T_{\rm IR}^{\rm dS}\left(\frac{\sigma}{F}\right)^{4}\cr\cr
&+&2V_{\rm inf}(\chi)\left(\frac{\sigma}{F}\right)^{2}\cr\cr
&-&|\alpha| c_{V}^{2}V_{\rm inf}(\chi)^{\frac{3}{4}}k+|\alpha| c_{\rm IR}^{2}k^{4}\left(\frac{\sigma}{F}\right)^{4}\cr\cr
&+&2(2+\alpha)k^{4}\left(\frac{\sigma}{F}\right)^{4}\left[c_{\rm IR}-c_{V}c_{\rm inf}\left(\frac{\sigma}{F}\right)^{\alpha}\right]^{2},\,\nonumber\\ 
\label{eq:V4Deff}
\ea
where the parametrization \eqref{eq:parametersink} of $V_{\rm inf}(\chi)$ has been used in the final line. The first line arises from the UV and IR brane-tension detunings, following from (\ref{eq:V4Dscalarpart}). The second line originates from the non-minimal coupling of the radion $\sigma$ to the 4D Ricci scalar $R^{(4)}$~\cite{Goldberger:1999un}
\be
\mathcal{L}_{\rm 4D}\supset-\frac{M_{4}^{2}}{2}R^{(4)}\left(\frac{\sigma}{F}\right)^{2}=-6M_{4}^{2}H_{\rm inf}^{2}\left(\frac{\sigma}{F}\right)^{2}\,,
\ee
where $R^{(4)}=12H_{\rm inf}^{2}$ for a 4D de Sitter background. The remaining terms follow from integrating $\Phi$ over the extra dimension in (\ref{eq:SGWchi}). 

To obtain an inflationary 4D de Sitter vacuum with a stabilized radion at
$\sigma_{\rm min}\neq0$ for $\alpha<0$, we first absorb the
$\sigma$-independent Goldberger-Wise contribution into the UV-localized vacuum
energy by choosing
\be
U_{\rm UV}(\chi)=|\alpha| c_{V}^{2}V_{\rm inf}(\chi)^{\frac{3}{4}}k\,.
\label{eq:VcancellationUV}
\ee
This tuning fixes the normalization of the inflationary vacuum energy but does
not affect the $\sigma$-dependent radion potential.
The remaining parameters are then chosen so that the radion sits at a stationary
minimum with vacuum energy
\be
V_{\rm 4D}(\chi,\sigma_{\rm min})=\xi V_{\rm inf}(\chi),\quad
V_{\rm 4D,\sigma}(\chi,\sigma_{\rm min})=0\,,
\label{eq:Vcancellation}
\ee
where $\xi>0$ and $V_{\rm 4D,\sigma}$ denotes differentiation with respect to
$\sigma$. The parameter $\xi$ measures the inflationary 4D vacuum energy in
units of $V_{\rm inf}(\chi)$ so that $H_{\rm inf}$ depends on $\xi$. In particular, for $\xi\simeq 1$, the Hubble scale is set
approximately by $V_{\rm inf}(\chi)$. With \eqref{eq:VcancellationUV} imposed, the effective 4D scalar potential becomes
\ba
&&V_{\rm eff}(\chi,\sigma)
=V_{\rm inf}(\chi)\cr\cr
&&+\,\delta T_{\rm IR}^{\rm dS}\left(\frac{\sigma}{F}\right)^{4}+2\xi V_{\rm inf}(\chi)\left(\frac{\sigma}{F}\right)^{2}+|\alpha| c_{\rm IR}^{2}k^{4}\left(\frac{\sigma}{F}\right)^{4}\cr\cr
&&+\,2(2+\alpha)k^{4}\left(\frac{\sigma}{F}\right)^{4}\left[c_{\rm IR}-c_{V}c_{\rm inf}\left(\frac{\sigma}{F}\right)^{\alpha}\right]^{2}\,,
\label{eq:Vsigma2}
\ea
For fixed $\alpha, c_V$ and $\xi$, the parameters $c_{\rm inf}$ and $c_{\rm IR}$ are then chosen so that this potential has the desired inflationary minimum. 

After reheating, the UV-brane contributions generated by the inflaton sector, including $U_{\rm UV}(\chi)$ and the induced Goldberger-Wise boundary contribution canceled in \eqref{eq:VcancellationUV}, disappear as the inflaton decays to its minimum. By contrast, the IR-brane detuning $\delta T_{\rm IR}^{\rm dS}$ introduced to support the inflationary de Sitter slicing is a fixed brane-tension contribution and remains in the late-time radion potential. This term should not be canceled as a $\sigma$-dependent contribution, since it is part of the late-time radion stabilization. 

However, its value at the late-time minimum of the post-inflationary radion potential, $\sigma_0/F =e^{-k d_0}$, where $d_0$ denotes the late-time inter-brane separation, does contribute to the 4D cosmological constant. We assume that the usual late-time brane-tension tuning, denoted by $T_{\rm UV}^{(0)}$ in (\ref{eq:IsraelJunction}), cancels
\be
V_{\rm GW}(\sigma_0)+\delta T_{\rm IR}^{\rm dS} e^{-4k d_0}
\ee
up to the observed dark-energy density today, where $V_{\rm GW}$ and $\delta T_{\rm IR}^{\rm dS}$ are given in \eqref{eq:Vsigma}  and \eqref{eq:irdetuning}, respectively. Crucially, however, this additional brane-tension tuning does not qualitatively affect the radion dynamics, because it remains parametrically small compared to $-\Lambda_{5}/k$.

It is worth emphasizing that $\sigma_{\rm min}\neq0$ is guaranteed throughout the entire cosmological evolution thanks to $\delta T_{\rm IR}^{\rm dS}$ which was required for 5D inflation. In this respect, $\delta T_{\rm IR}^{\rm dS}$ accomplishes two goals simultaneously in the present framework: it is essential both for achieving the 5D inflationary background and for ensuring the persistence of radion stabilization at $\sigma_{\rm min}\neq0$ in both pre- and post-inflationary eras.

In the 4D effective theory described by (\ref{eq:Vsigma2}), only the inflaton $\chi$ and the radion $\sigma$ become relevant. Because $m_{\rm KK}\gg H_{\rm inf}$ during inflation, the spin-2 and Goldberger-Wise scalar Kaluza--Klein modes effectively decouple from the system. In \cite{Choi:2025lkg},  the MSSM matter fields were localized on the UV brane, while the gauge bosons propagate in the bulk. Within this framework, the inflaton can couple to both gauge field and matter zero modes on the UV brane, thereby giving rise to natural decay channels through which the inflaton reheats the universe.

Since the radion potential in (\ref{eq:Vsigma2}) inherits an explicit time dependence from $V_{\rm inf}(\chi)$, the radion minimum evolves accordingly as
\be
\sigma_{\rm min}=\begin{cases}
    Fe^{-kd_{\rm inf}},\qquad a<a_{\rm end}\,,\\
    F e^{-kd_{0}},\qquad~ a\gtrsim a_{c}\,,
\end{cases}
\label{eq:sigmamincases}
\ee
where $a_{\rm end}$ and $a_{c}$ denote the scale factors at the time when the inflation ends (i.e. $\ddot{a}=0$) and $c_{V}V_{\rm inf}(\chi)^{\frac{3}{8}}=v_{\rm UV}\equiv c_{\rm UV}k^{\frac{3}{2}}$, respectively. As the value of $V_{\rm inf}(\chi)$ barely changes during inflation, the crossover takes place after inflation ends, $a_{c}>a_{\rm end}$. 

Consequently, the enhanced UV boundary value in (\ref{eq:enhancedBV}) stabilizes the radion during inflation at a smaller inter-brane separation $d_{\rm inf}$ than in the post-inflationary Goldberger–Wise minimum $d_{0}$, which is recovered at ($a\gtrsim a_{c}$). The value of $\sigma_{\rm min}$ is determined in each cosmological era by minimizing the $\sigma$-dependent part in (\ref{eq:Vsigma2}). Thus, the system begins inflation with a relatively small brane separation.
After inflation ends, the UV boundary value relaxes to $v_{\rm UV}$, with the subsequent increase in the inter-brane separation controlled by the relative size of $c_{V} V_{\rm inf}(\chi)^{3/8}$, compared with $v_{\rm UV}$.

In the dual 4D description, the radion is the dilaton associated with spontaneous breaking of approximate conformal symmetry, and the IR brane position sets the CFT confinement scale, $
\Lambda_{\rm CFT}\sim k e^{-kd}$.
The Goldberger-Wise scalar corresponds to a relevant CFT deformation with operator dimension
$\Delta=2+2\sqrt{1+\alpha}$,
which for the benchmark $\alpha=-0.3$ gives $\Delta\simeq 3.67$. The inflaton is an elementary UV-sector field, dual to the UV-brane-localized scalar $\chi$, and its coupling to the Goldberger-Wise sector appears in the 4D dual as a coupling to the dilaton potential. During inflation this coupling temporarily raises the confinement scale, $\Lambda_{\rm CFT}^{\rm inf}>\Lambda_{\rm CFT}^{0}$,
and hence enhances the axion decay constant,
$f_{\rm inf}\sim \sqrt{N_{\rm CFT}}\Lambda_{\rm CFT}^{\rm inf}>
f_a\sim \sqrt{N_{\rm CFT}}\Lambda_{\rm CFT}^{0}$.
After reheating, the inflaton-induced source disappears, the confinement scale relaxes to its late-time value, and the axion decay constant returns to the QCD axion window.

Since the radion couples to the inflaton via the interaction in (\ref{eq:Vsigma2}) from $c_{\rm inf}\propto V_{\rm inf}(\chi)^{3/8}$, the scalar fluctuation spectrum depends on the background value of $\chi$, inducing a radiative correction to the inflaton potential~\cite{Coleman:1973jx}. During inflation, the radion obtains an effective mass-squared
\be
\label{eq:msigmaeff2}
m_{\sigma,{\rm eff}}^{2}(\chi)=\left.\frac{\partial^{2}V_{\rm eff}}{\partial\sigma^{2}}\right|_{\sigma=\sigma_{\rm min}}\,.
\ee
The corresponding Coleman-Weinberg one-loop contribution\footnote{Although $\delta V_{\rm inf}$ exhibits only logarithmic sensitivity in $\Lambda_{\rm UV}=m_{\rm KK}$ when computed using dimensional regularization, a quadratic divergence of the form $\delta V_{\rm inf}\propto \Lambda_{\rm UV}^{2}$ appears once an explicit ultraviolet cutoff is imposed (or, equivalently, when divergences are regulated via Pauli–Villars fields). In this work, we assume that such quadratic contributions are canceled by a counterpart contribution, as can naturally occur within a supersymmetric framework. Under this assumption, the residual effect after partial cancellation yields $\delta V_{\rm inf}(\chi)\propto m_{3/2}^{2}$, where the gravitino mass $m_{3/2}$ satisfies $m_{3/2}\ll H_{\rm inf}$.}, evaluated at the renormalization scale $\mu=m_{\rm KK}$, is\footnote{There is also an interaction between $V_{\rm inf}(\chi)$ and the scalar perturbation $S(x,y)$ of the AdS$_{5}$ metric on the UV brane~\cite{Goldberger:1999un,Creminelli:2001th,Csaki:2000zn,Csaki:2007ns}
\be
ds^{2}=e^{-2ky-2S}\eta_{\mu\nu}dx^{\mu}dx^{\nu}-(1+2S)dy^{2}\,,
\ee
where $S(x,y)=(\frac{s(x)}{F}e^{-ky_{\rm IR}})\,e^{2ky}$ with $s(x)$ the canonically normalized 4D scalar field. The fluctuation $\delta r(x)$ of $r(x)$ at $y$ is related to $S(x,y)$ via $\pi k\delta r(x)=S(x,y)$. The coupling of $S(x,y)$ to $\chi$ on the UV brane induces only a negligible radiative correction to $\delta V_{\rm inf}(\chi)$ via $s(x)$ loops, of order $\delta V_{\rm inf}(\chi)=\mathcal{O}(e^{-4ky_{\rm IR}}H_{\rm inf}^{8}/F^{4})\ll\mathcal{O}(H_{\rm inf}^{4})$ during inflation.}
\be
\delta V_{\rm inf}(\chi)\approx\frac{m_{\sigma,{\rm eff}}^{4}(\chi)}{64\pi^{2}}\log\left[\frac{m_{\sigma,{\rm eff}}^{2}(\chi)}{m_{\rm KK}^{2}}\right]\,.
\label{eq:deltaVchi}
\ee
Crucially, $V_{\rm inf}(\chi)$ and $\sigma_{\rm min}$ are approximately constant over the $\chi$-field range traversed during inflation, so $m_{\sigma,{\rm eff}}^{2}(\chi)$ in (\ref{eq:msigmaeff2}) is also constant throughout the inflationary epoch. 
Consequently, $\delta V_{\rm inf}(\chi)$ in (\ref{eq:deltaVchi}) has negligible $\chi$-dependence, amounting primarily to a small renormalization of the overall height of the inflaton potential. Since $\delta V_{\rm inf}(\chi)=\mathcal{O}\left(m_{\sigma,{\rm eff}}^{4}(\chi)\right)$, this correction is at most of order $\mathcal{O}((10H_{\rm inf})^{4})$ for the parameters we consider below, and is therefore parametrically suppressed relative to the tree-level energy density $V_{\rm inf}(\chi)=3M_{4}^{2}H_{\rm inf}^{2}$.

\subsection{Consistency Conditions and Parameter Constraints}
For the scenario described by (\ref{eq:sigmamincases}) to be phenomenologically viable, the parameter space must satisfy two essential requirements. First, if $d_{\rm inf}$ is to remain constant during inflation, $\sigma$ must sit at $\sigma_{\rm min}$ and remain stationary throughout the inflationary epoch $a<a_{\rm end}$. Second, the additional interactions introduced in (\ref{eq:SGW2}) and (\ref{eq:Vsigma2}) must not spoil the inflationary dynamics. These requirements lead to the following consistency conditions:
\begin{itemize}
    \item {\bf C1}: $\sigma$ stays at the transient minimum of the $\sigma$-dependent part of $V_{\rm eff}(\chi,\sigma)$ in (\ref{eq:Vsigma2}) during inflation, with an effective mass $m_{\sigma,{\rm eff}}$ in (\ref{eq:msigmaeff2}) larger than or comparable to the Hubble expansion rate    
\be
m_{\sigma,{\rm eff}}\gtrsim H_{\rm inf},\quad\text{for $a<a_{\rm end}.$}
\label{eq:heavysigma}
\ee
\item {\bf C2}: Assuming that the inflationary model defined by $V_{\rm inf}(\chi)$ is consistent with the observed spectral-index constraint $0.956<n_{s}<0.973$ (95\% C.L.)~\cite{Balkenhol:2025wms}, the radion-induced contributions to the slow-roll parameters must be sufficiently suppressed
\ba
|\delta\epsilon_{\chi,0}|&=&|\epsilon_{\rm inf}-\epsilon_{\chi,0}|\ll1\,,\nonumber\\
%\cr\cr\quad
|\delta\eta_{\chi,0}|&=&|\eta_{\rm inf}-\eta_{\chi,0}|\ll1\,.
\label{eq:deltaepeta}
\ea
\end{itemize}
The slow-roll parameters in {\bf C2} derived solely from the inflaton potential $V_{\rm inf}(\chi)$ are written as
\be
\epsilon_{\chi,0}=\frac{M_{4}^{2}}{2}\left(\frac{V_{\rm inf,\chi}}{V_{\rm inf}}\right)^{2}\ll 1,\quad
\eta_{\chi,0}=M_{4}^{2}\frac{V_{\rm inf,\chi\chi}}{V_{\rm inf}}\ll 1\,,
\label{eq:slowroll0}
\ee
while those derived from $V_{\rm eff}(\chi,\sigma)$ in (\ref{eq:Vsigma2}) are    
\be
\epsilon_{\rm inf}=\frac{M_{4}^{2}}{2}
\left(\frac{V_{\rm eff,\chi}}{V_{\rm eff}}\right)^{2},\quad\eta_{\rm inf}=\sum_{I,J=1}^{2}n^{I}n^{J}\eta_{IJ}\,,
\label{eq:slowroll}
\ee
with $\vec{n}$ and $\eta_{IJ}$ defined as
\be
\vec{n}=(\cos\theta,\sin\theta),\quad\eta_{IJ}\equiv \frac{M_{4}^{2}}{V_{\rm eff}}\begin{pmatrix}
V_{\rm eff,\chi\chi} & V_{\rm eff,\chi\sigma}\\
V_{\rm eff,\sigma\chi} & V_{\rm eff,\sigma\sigma}
\end{pmatrix}\,,
\ee
and $\theta$ denotes the polar angle in the two-dimensional field space spanned by $(\chi,\sigma)$, given by
\be
\tan\theta=\frac{\dot{\sigma}}{\dot{\chi}}\simeq-\frac{V_{\rm eff,\sigma\chi}}{V_{\rm eff,\sigma\sigma}}\,.
\label{eq:tantheta}
\ee 
 The subscript on $V_{{\rm eff},A_{1}...A_{n}}$ indicates the derivatives with respect to the fields $A_{i}=\chi,\sigma$ and all quantities are evaluated at $\sigma=\sigma_{\rm min}$ given in (\ref{eq:sigmamincases}) for $a<a_{\rm end}$. For the heavy $\sigma$ satisfying {\bf C1}, the second equality in (\ref{eq:tantheta}) can be derived by differentiating $V_{\rm eff,\sigma}(\chi,\sigma_{\rm min}(\chi))=0$ with respect to time, which yields
\be
V_{\rm eff,\sigma\sigma}\dot{\sigma}+V_{\rm eff,\sigma\chi}\dot{\chi}=0\,.
\ee

For $\dot{\sigma}\ll\dot{\chi}$, $\tan\theta\simeq\sin\theta$ and therefore $\eta_{\rm inf}$ evaluated at $\sigma=\sigma_{\rm min}$ in (\ref{eq:sigmamincases}) becomes
\be
\eta_{\rm inf}=\eta_{11}\cos^{2}\theta+2\eta_{12}\sin\theta\cos\theta+\eta_{22}\sin^{2}\theta\,.
\label{eq:etainf}
\ee
The parameter $\epsilon_{\chi,0}$ is related to $H_{\rm inf}$ (and thus $c_{\rm inf}$) via
\be
\epsilon_{\chi,0}=\frac{H_{\rm inf}^{2}}{8\pi^{2}M_{4}^{2}A_{s}}\simeq\frac{c_{\rm inf}^{\frac{8}{3}}}{3b_{\rm CFT}^{4}}\left(\frac{M_{4}}{10^{15}{\rm GeV}}\right)^{2}\lesssim\mathcal{O}(10^{-3})\,,
\ee
where $A_{s}=2\times10^{-9}$ is the amplitude of the adiabatic perturbation power spectrum at the CMB pivot scale $k_{\rm CMB}=0.05\,{\rm Mpc}^{-1}$ and $\eta_{\chi,0}$ is typically given by $\mathcal{O}(10^{-2})$~\cite{Balkenhol:2025wms}. Hence, for an inflation model consistent with the observed scalar spectral index $n_{s}$, where the latter is determined by the slow-roll parameters, $\epsilon_{\chi,0}$ and $\eta_{\chi,0}$, through $n_{s}=1-6\epsilon_{\chi,0}+2\eta_{\chi,0}$, the condition \eqref{eq:deltaepeta} in {\bf C2} translates into
\ba
{\bf C2}:\,\,\,|\delta\epsilon_{\chi,0}|&=&|\epsilon_{\rm inf}-\epsilon_{\chi,0}|<|\epsilon_{\chi,0}|\,,\nonumber\\
|\delta\eta_{\chi,0}|&=&|\eta_{\rm inf}-\eta_{\chi,0}|<|\eta_{\chi,0}|\sim10^{-2}\,.
\label{eq:cinfmax2}
\ea

%%%%%%%%%%%%%%%%%%%%%%%%%%%%%%%%%%%%%%%%%%%
\begin{figure}[t]
\centering
\hspace*{-5mm}
\includegraphics[width=0.49\textwidth]{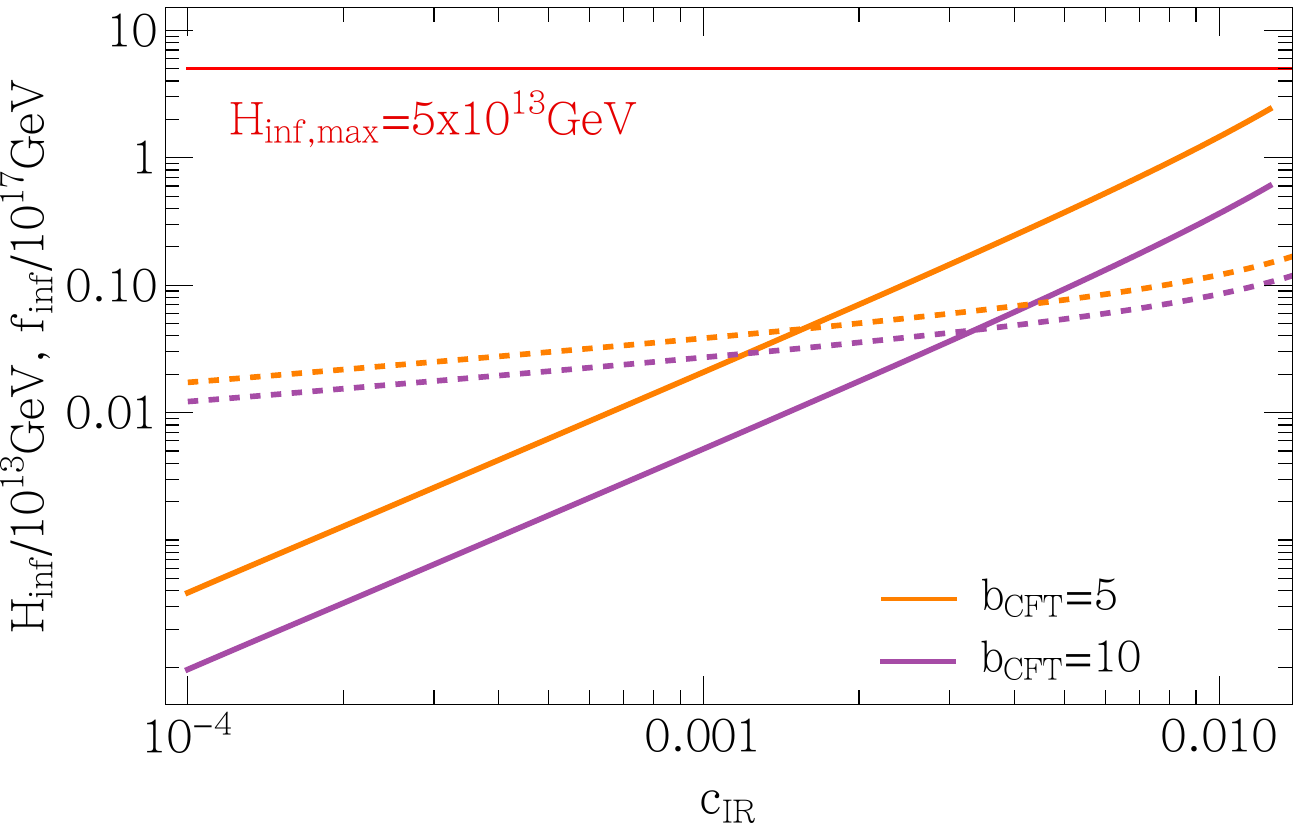}
\caption{The inflationary Hubble scale $H_{\rm inf}$ (solid) and effective axion decay constant $f_{\rm inf}$ (dashed) as functions of $c_{\rm IR}$ for $\alpha = -0.3$, $c_{V}=1$ and $\xi=0.99$. The {\bf{\color{orange}orange}} and {\bf{\color{violet} purple}} lines correspond to $ {\bf{\color{orange}b_{\rm CFT} =5}}$ and $ {\bf{\color{violet} b_{\rm CFT} =10}}$, respectively. The {\bf{\color{red} red}} line shows the observational upper bound, {\bf{\color{red} $H_{\rm inf} \le 5 \times 10^{13}\,\mathrm{GeV}$}}.}
\vspace*{-1.5mm}
\label{fig:HIFI}
\end{figure}
%%%%%%%%%%%%%%%%%%%%%%%%%%%%%%%%%%%%%%%%%%%

For a fixed set $(\alpha,c_{V},\xi)$ we numerically determine $(c_{\rm inf},\sigma_{\rm min}/F)$ that satisfy (\ref{eq:Vcancellation}) for each value of $c_{\rm IR}$. These solutions allow us to evaluate $f_{\rm inf}$ and $H_{\rm inf}$ using (\ref{eq:fa}) and (\ref{eq:parametersink}) respectively. In the following analysis, we adopt the benchmark choice $\alpha=-0.3$, $c_{V}=1$ and $\xi=0.99$ with $H_{\rm inf}\simeq \sqrt{V_{\rm inf}(\chi)/(3M_{4}^{2})}$.

The resulting behavior is shown in Fig.~\ref{fig:HIFI}, where $H_{\rm inf}$ and $f_{\rm inf}$ are displayed as functions of $c_{\rm IR}$ for two representative values of $b_{\rm CFT}$. For $c_{\rm IR}\lesssim 0.01$ and $b_{\rm CFT}=5,10$, the consistency conditions {\bf C1} and {\bf C2} can be simultaneously satisfied even for a high inflationary scale approaching the current observational upper bound (red solid) $H_{\rm inf}=\pi M_{4}\sqrt{\mathcal{P}_{t}(k_{\rm CMB})/2}\simeq5\times10^{13}$~GeV, inferred from the constraint on the primordial tensor power spectrum. 

For $c_{\rm IR}\lesssim0.01$, we find that $\sigma_{\rm min}/F$ can reach values as large as $\sim0.14$ at $c_{\rm IR}=0.01$, and decreases monotonically as $c_{\rm IR}$ is lowered. The effective radion mass in (\ref{eq:msigmaeff2}) was found to satisfy $m_{\rm \sigma,eff}\gg H$, ensuring the validity of {\bf C1}. Furthermore, we numerically found that the modifications to the slow-roll parameters remain moderate, with 
\be
\epsilon_{\rm inf}=(0.8-0.9)\,\epsilon_{\chi,0}\,,\quad\eta_{\rm inf}=(0.9-0.95)\,\eta_{\chi,0}\,,
\ee
satisfying {\bf C2}. Consequently, our framework accommodates a broad class of inflationary models, provided that the resulting scalar spectral index
\be
n_{s}=1-6\epsilon_{\rm inf}+2\eta_{\rm inf}\,,
\ee
is consistent with the observed value $n_{s}=0.965\pm0.004$ (68\% C.L.)~\cite{Balkenhol:2025wms}. 

Notably, for $c_{\rm IR}\lesssim0.01$, the effective decay constant during inflation can be as large as $f_{\rm inf}=\mathcal{O}(10^{16})\,{\rm GeV}$. This enhancement follows from the smaller inter-brane separation $d_{\rm inf}< d_{0}$ realized during inflation, before the radion evolves to its post-inflationary value at $d_{0}$. 

%%%%%%%%%%%%%%%%%%%%%%%%%%%%%%%%%
\begin{figure*}[ht]
  \centering
  \hspace*{-5mm}
  \subfigure{\includegraphics[scale=0.411]{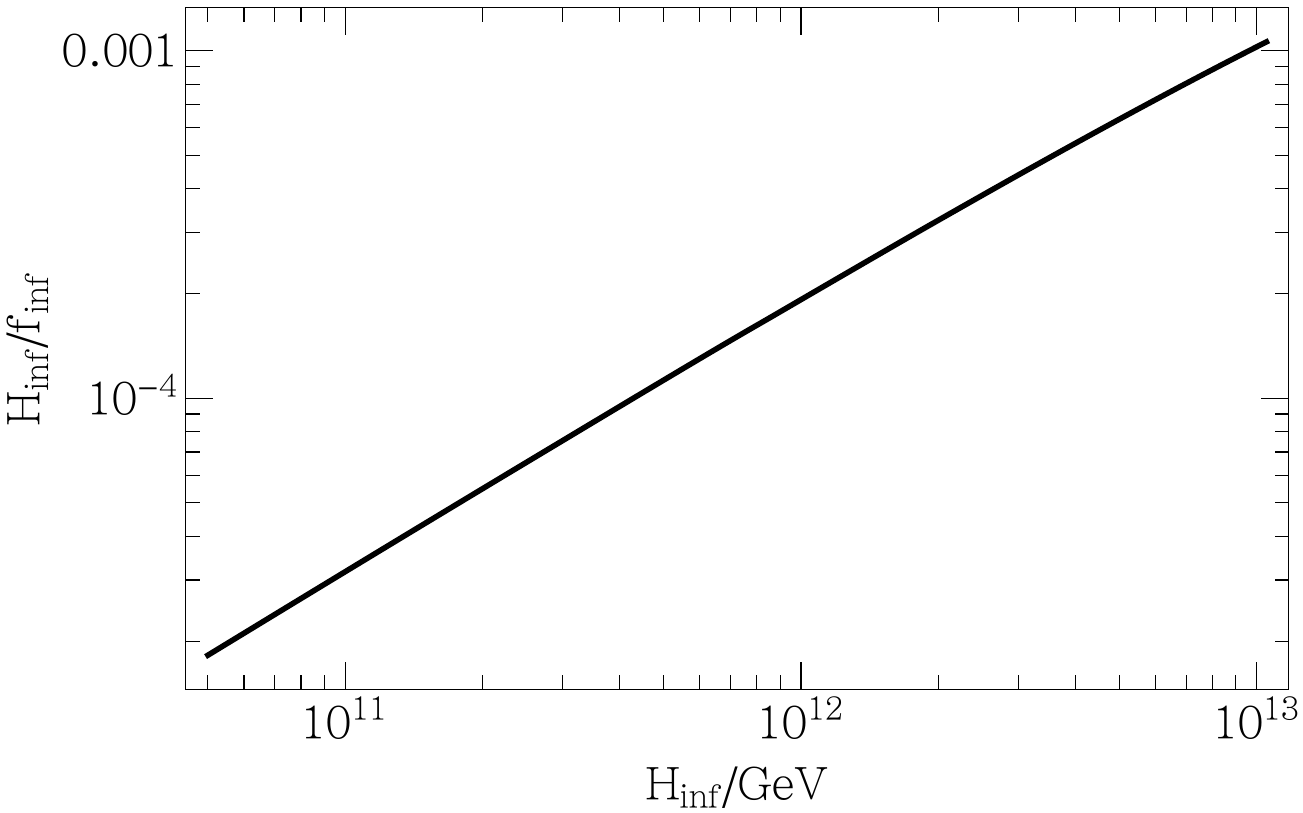}}
  \subfigure{\includegraphics[scale=0.411]{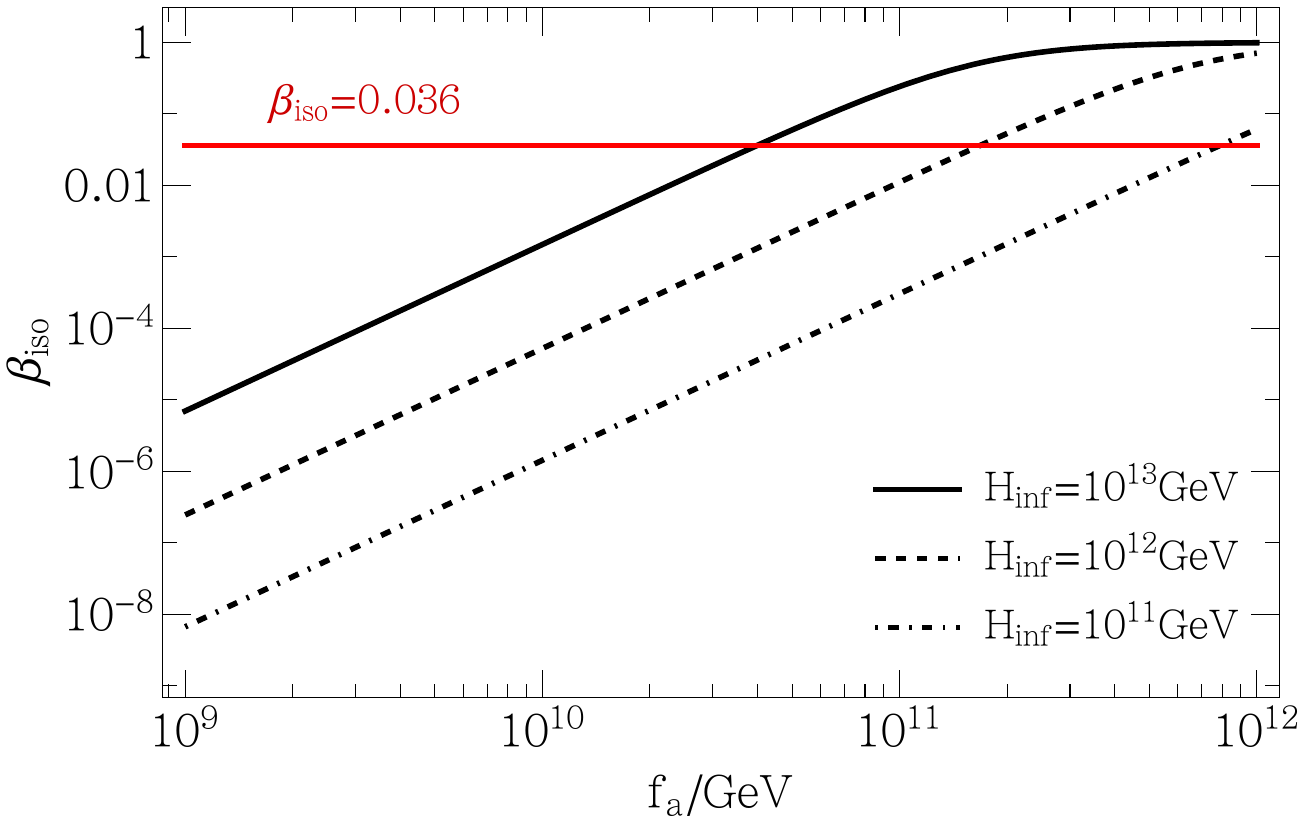}}
  \caption{ {\bf Left}: The ratio $H_{\rm inf}/f_{\rm inf}$ as a function of $H_{\rm inf}$ for $\alpha = -0.3$, $c_{V}=1$ and $\xi=0.99$. {\bf Right}: The isocurvature fraction $\beta_{\rm iso}$ as a function of the late-time axion decay constant $f_{a}$. The solid, dashed and dot-dashed lines correspond to $H_{\rm inf} = 10^{13}\mathrm{GeV}$, $10^{12}{\rm GeV}$ and $10^{11}\mathrm{GeV}$, respectively. The {\bf{\color{red} red}} horizontal line shows the observational upper bound $ {\bf{\color{red} \beta_{\rm iso}=0.036}}$.}
  \vspace*{-1.5mm}
\label{fig:ratiobetaiso}
\end{figure*}
%%%%%%%%%%%%%%%%%%%%%%%%%%%%%%%%%

We further find that decreasing $|\alpha|$ leads to larger values of both $c_{\rm inf}$ and $\sigma_{\rm min}/F$ while simultaneously increasing the shifts $\delta \epsilon_{\chi,0}$ and $\delta\eta_{\chi,0}$. In the next section, we examine how the choice of $\alpha$ affects the axion isocurvature perturbation.

%%%%%%%%%%%%%%%%%%%%%%%%%%%%%%%%%%%%%%%%%%

\section{Suppressing axion isocurvature}
\label{sec:isocurvature}

In Sec.~\ref{sec:smallextraD}, we showed how the inflationary Hubble scale, $H_{\rm inf}$, and the effective axion decay constant during inflation, $f_{\rm inf}$, are governed by the microscopic parameters $(\alpha,c_{\rm IR},c_{\rm inf},c_{V},b_{\rm CFT})$ appearing in (\ref{eq:fa}) and (\ref{eq:parametersink}). These quantities collectively control the inflationary dynamics and cause the transient modification of the size of the extra dimension. By contrast, the late-time axion decay constant $f_{a}$ (for $a \gtrsim a_{c}$) is determined by UV brane value $v_{\rm UV}$, which—together with $v_{\rm IR}$ and $\alpha$—fixes the present-day warp factor and hence the low-energy normalization of the axion field. In this section, we investigate how a time-dependent axion decay constant, induced by the dynamical evolution of the size of the extra dimension, can suppress axion isocurvature perturbations at CMB scales.

Relative to the conventional scenario with $f_{\rm inf}=f_{a}$, the isocurvature power spectrum $\mathcal{P}_{\rm iso}(k_{\rm CMB})$ in (\ref{eq:Piso}) acquires an additional parametric suppression by a factor of $(f_{a}/f_{\rm inf})^{2}$ whenever $f_{\rm inf}\gg f_{a}$. Physically, the enhancement of the decay constant during inflation helps to further suppress the angular fluctuation $\delta\theta_{a,i}$ at horizon crossing, thereby attenuating the resulting isocurvature amplitude.

In the case where $f_{\rm inf} \neq f_{a}$, the expression in (\ref{eq:Piso}) should be regarded as a function of four independent parameters: $f_{\rm inf}$, $f_{a}$, the initial misalignment angle $\theta_{a,i}$, and the inflationary Hubble scale $H_{\rm inf}$. To adopt a conservative benchmark, we fix the initial angle to $\theta_{a,i}=1$, which avoids any artificial suppression arising from a tuned initial condition.

Independent of the post-inflationary decay constant $f_{a}$ (which is set by the choice of $v_{\rm UV}$), one can search for values of $(\alpha,\xi, c_{\rm IR},c_{V},c_{\rm inf},e^{-kd_{\rm inf}})$ by solving (\ref{eq:Vcancellation}) and check if it is consistent with {\bf C1} and {\bf C2} as discussed in Sec.~\ref{sec:smallextraD}. Consequently, for a given value of $b_{\rm CFT}$, a set of parameter values for $(\alpha,\xi, c_{\rm IR},c_{V},c_{\rm inf},e^{-kd_{\rm inf}})$ uniquely fixes the corresponding inflationary scales $H_{\rm inf}$ and $f_{\rm inf}$ that satisfy the requirements {\bf C1} and {\bf C2} (see Fig.~\ref{fig:HIFI}). In this way, for fixed $\alpha, c_{V}$ and $\xi$, one can scan over $c_{\rm IR}$ to evaluate the ratio $H_{\rm inf}/f_{\rm inf}$ as well as the isocurvature perturbation fraction $\beta_{\rm iso}$ within the canonical QCD axion window $10^{9}\,{\rm GeV} \lesssim f_{a} \lesssim 10^{12}\,{\rm GeV}$.

In Fig.~\ref{fig:ratiobetaiso}, we present the results for $\alpha = -0.3$, $c_{V}=1$ and $\xi=0.99$. The left panel shows the ratio $H_{\rm inf}/f_{\rm inf}$ as a function of $H_{\rm inf}$, imposing the consistency conditions {\bf C1} and {\bf C2}. The difference between $b_{\rm CFT}=5$ and $10$ is not distinguishable in the figure. The allowed values are obtained by scanning over $10^{-4} \lesssim c_{\rm IR} \lesssim 10^{-2}$ and using the corresponding $H_{\rm inf}$ and $f_{\rm inf}$ values from Fig.~\ref{fig:HIFI}. At fixed $c_{\rm inf}$ and $d_{\rm inf}$, the ratio depends on $b_{\rm CFT}$, obeying the scaling $H_{\rm inf}/f_{\rm inf}\propto b_{\rm CFT}^{-\frac3{2}}$.

We then use the axion relic abundance~\cite{ParticleDataGroup:2024cfk} 
\be
\Omega_{a}h^{2} \simeq 0.12\,\theta_{a,i}^{2} \left(\frac{f_{a}}{9\times10^{11}\,{\rm GeV}}\right)^{1.165},
\ee
where the observed cold dark matter density $\Omega_{\rm cdm}h^{2}\simeq 0.12$ with $h\simeq 0.67$ 
to compute the isocurvature perturbation fraction $\beta_{\rm iso}$ defined in (\ref{eq:betaiso}). For each value of $H_{\rm inf}$, we use the corresponding ratio $H_{\rm inf}/f_{\rm inf}$, assuming the absence of possible entropy dilution of the axion relic abundance before BBN, as well as anharmonic corrections to the axion potential.

For high-scale inflation with $H_{\rm inf} \simeq 10^{13}\,{\rm GeV}$, the predicted isocurvature fraction $\beta_{\rm iso}$ exceeds the observational upper bound for $f_{a} \gtrsim 4\times 10^{10}\,{\rm GeV}$. More generally, if the axion zero mode constitutes all of the dark matter, our framework does not allow $H_{\rm inf}\sim10^{13}{\rm GeV}$ and $f_{a}\sim10^{12}{\rm GeV}$ compatible with $\beta_{\rm iso}\leq0.036$ unless the axion relic abundance is diluted or the initial misalignment angle satisfies $\theta_{a,i}\ll1$. 

For lower inflationary scales, however, the mechanism can substantially relax the isocurvature constraint.
In particular, for $H_{\rm inf} \simeq 10^{12}\,{\rm GeV}$, the predicted axion isocurvature fraction satisfies  $\beta_{\rm iso} < 0.036$ for $f_{a}$ as large as $2\times10^{11}\,{\rm GeV}$ without relic abundance dilution or a fine-tuned small $\theta_{a,i}$. The enhancement of $f_{\rm inf}$ therefore reopens a broad region of parameter space that would otherwise be excluded by the observational isocurvature bound. For $H_{\rm inf} \simeq 10^{12}\,{\rm GeV}$, $\beta_{\rm iso}$ can be compatible with observations if the coherent axion zero mode is a subdominant dark matter component, contributing less than $\sim 20\%$ of the total dark matter density in the standard cosmology. Of course, when our mechanism is realized together with the relic abundance dilution or a fine-tuned small $\theta_{a,i}$, the fraction of the dark matter contributed by the axion can be larger.
 
Varying $\alpha$ and $\xi$ does not significantly change this conclusion. For $|\alpha|<0.3$ or $\xi<0.99$,
the solutions of (\ref{eq:Vcancellation}) yield 
larger values of both $H_{\rm inf}$ and $f_{\rm inf}$, but their ratio remains close to that of the benchmark point $|\alpha|=0.3$, $\xi=0.99$.
Thus, changing $|\alpha|$ and $\xi$ does not necessarily yield a smaller $\beta_{\rm iso}$. Moreover, the induced shifts in the slow-roll parameters become more pronounced for $|\alpha|<0.3$ and $\xi<0.99$, imposing additional restrictions on the inflationary models compatible with the mechanism.

%%%%%%%%%%%%%%%%%%%%%%%%%%%%%%%%%
\section{Small-Scale Fluctuations and Domain-Wall Avoidance}
\label{sec:smallscale}

Although a time-dependent axion decay constant is useful for suppressing the large-scale fluctuation $\delta\theta_a$, a rapid decrease from $f_{\rm inf}$ to $f_a$ can enhance fluctuations in the angular velocity $\dot{\theta}_{a}$ even on superhorizon scales. This effect can subsequently amplify 
axion  perturbations on small scales. As a result, the isocurvature power spectrum is enhanced for modes that re-enter the horizon during the interval $a_{f_{\rm inf}} < a < a_{f_a}$, where $a_{f_{\rm inf}}$ and $a_{f_a}$ denote the onset and completion of the evolution of the axion decay constant, respectively.

When the axion decay constant evolves as $f\propto a^{-n}$ during the interval $a_{f_{\rm inf}}\lesssim a<a_{f_a}$, the superhorizon angular fluctuation modes remain constant at leading order before the transition begins, $a\lesssim a_{f_{\rm inf}}$~\cite{Kobayashi:2016qld}. However, once $f$ begins to decrease, a growing superhorizon mode appears in $\delta\theta_{a,k}$ for modes with wavenumber larger than
\be
k_{*}= \sqrt{2n-\frac{3}{2}} \,(aH)\left(\frac{f}{f_{\rm inf}}\right)^{1-\frac{5}{4n}}\, .
\ee
For $n>5/4$, the characteristic scale $k_*(a)$ is initially of order the horizon scale at the onset of the transition, $k_*(a_{f_{\rm inf}})\sim a_{f_{\rm inf}}H(a_{f_{\rm inf}})$. As $f$ decreases, $k_*$ moves toward smaller comoving wavenumbers, or larger physical length scales, eventually reaching
\be
k_*(a_{f_a})=\sqrt{2n-\frac{3}{2}} a_{f_a}H(a_{f_a})\left(\frac{f_a}{f_{\rm inf}}\right)^{1-\frac{5}{4n}}\,,
\ee
by the end of the transition. Consequently, $k_*(a_{f_a})$ sets the smallest comoving wavenumber, or equivalently largest wavelength, for which $\delta\theta_{a,k}$ is dominated by the growing superhorizon mode.

To check that this scale is far below CMB scales, we compare it with a
representative CMB mode $k_{\rm CMB}$ and estimate
\ba
\frac{k_{*}(a_{f_a})}{k_{\rm CMB}}&=&\frac{a_{f_a}H(a_{f_a})}{a_{\rm CMB}H(a_{\rm CMB})}\times \sqrt{2n-\frac{3}{2}}\left(\frac{f_{a}}{f_{\rm inf}}\right)^{1-\frac{5}{4n}}\cr\cr
&\simeq&\frac{a_{\rm eq}}{a_{f_a}}\sqrt{2n-\frac{3}{2}}\left(\frac{f_{a}}{f_{\rm inf}}\right)^{1-\frac{5}{4n}}\,,
\label{eq:kscompare}
\ea
where $a_{\rm eq}$ is the scale factor at matter--radiation equality. To obtain the second line, we used $aH\propto a^{-1}\,(a^{-1/2})$ during the radiation- (matter-) dominated epoch, where the factor $\sqrt{a_{\rm CMB}/a_{\rm eq}}\simeq1.76$ is omitted. Eq.~(\ref{eq:kscompare}) therefore implies that $k_{*}(a_{f_a})>k_{\rm CMB}$ 
by the factor shown in the second line, and hence the corresponding physical length scale is smaller than the CMB scale.

As an illustrative example, suppose that the axion decay constant settles to its present value $f_a$ when the primordial plasma temperature $T\simeq 10^{9}\,\mathrm{GeV}$. For $n\gtrsim5/4$, the CMB length scale then exceeds the length scale associated with $k_{*}(a_{f_a})$ by at least the factor
\be
\frac{a_{\rm eq}}{a_{f_a}}\frac{f_a}{f_{\rm inf}}\simeq10^{18}\,\frac{f_a}{f_{\rm inf}}\,,
\label{eq:smalllargescale}
\ee
where $a_{\rm eq}$ corresponds to a temperature $T_{\rm eq}\simeq 1\,\mathrm{eV}$. Thus, for $f_{\rm inf}\lesssim 10^{16}{\rm GeV}$ and $f_{a}\gtrsim10^{9}\,{\rm GeV}$, the factor in (\ref{eq:smalllargescale}) exceeds $10^{11}$. The enhancement of the isocurvature spectrum therefore occurs only on scales far smaller than those probed by the CMB.

For modes that re-enter the horizon at $a=a_{f_a}$, defined by $k_{f_a}\equiv a_{f_a}H(a_{f_a})$, the isocurvature power spectrum for $a\gtrsim a_{f_a}$ is approximately~\cite{Kobayashi:2016qld} 
\be
\mathcal{P}_{\rm iso}^{\frac{1}{2}}(a,k_{f_a})\simeq\frac{H_{\rm inf}}{2\pi f_{a}}\left(\frac{f_{\rm inf}}{f_{a}}\right)^{1-\frac{5}{2n}}\frac{a_{f_a}}{a}\,.
\label{eq:Pisohalf}
\ee
This expression shows that for high-scale inflation with $H_{\rm inf}\gtrsim 10^{12}\,\mathrm{GeV}$, a large hierarchy $f_{\rm inf}\gg f_{a}$, and $n>5/2$, modes with $k\sim k_{f_a}$ can acquire sizeable small-scale isocurvature fluctuations after the transition from $f_{\rm inf}$ to $f_{a}$ is completed. If, by the time $a=a_{m}$ when the non-zero axion potential gets generated, the amplitude satisfies $\mathcal{P}_{\rm iso}^{\frac{1}{2}}(a_{m},k_{f_a})\gg 2\pi$, then the resulting inhomogeneous axion field characterized by $\mathcal{P}_{\rm iso}^{\frac{1}{2}}(a_{m},k_{f_a})$ within a horizon can trigger the formation of dangerous domain walls. Consequently, a viable time-dependent decay constant solution to the axion isocurvature problem must also satisfy
\be
\mathcal{P}_{\rm iso}^{\frac{1}{2}}(a_{m},k_{f_a})\ll 2\pi\,,
\ee
in order to avoid additional complications on small scales.

In our scenario, the axion decay constant begins evolving from $f_{\rm inf}$ to $f_{a}$ immediately after the end of inflation, so that $a_{f_{\rm inf}}=a_{\rm end}$. Because the transition to the present decay constant $f_a$ is most likely realized at $a\simeq a_{c}$, it is completed no later than reheating, i.e. $a_{\rm RH}>a_{f_a}$. Thus, one may conservatively replace $a_{f_a}$ with $a_{\rm RH}$ in (\ref{eq:Pisohalf}) to estimate $\mathcal{P}_{\rm iso}^{1/2}(a_{m},k_{f_a})$. Assuming that non-perturbative QCD effects provide the only contribution to the axion potential, then the axion mass turns on around $T\simeq \Lambda_{\rm QCD}$ and the ratio of scale factors satisfies
\be
\frac{a_{f_a}}{a_{m}}\lesssim \frac{a_{\rm RH}}{a_{m}}\simeq \frac{\Lambda_{\rm QCD}}{T_{\rm RH}}\,,
\label{eq:amaf}
\ee
where $\Lambda_{\rm QCD}\simeq 200\,\mathrm{MeV}$.

On the other hand, since $f\propto\sigma/F$, the scaling exponent $n$ is determined by  the radion field evolution $\sigma\propto a^{-n}$ during the interval $a_{f_{\rm inf}}<a<a_{f_a}$. As $\sigma$ decreases from $\sigma(a_{\rm end})$ to $\sigma(a_{c})$ respectively, the warp factor is reduced by a factor of $(a_{c}/a_{\rm end})^{n}$. Comparing the approximate expressions for the warp factors i.e. $(c_{\rm inf}/c_{\rm IR})^{\frac{1}{|\alpha|}}$ at $a=a_{\rm end}$, and $(c_{\rm UV}/c_{\rm IR})^{\frac{1}{|\alpha|}}$ at $a_{c}$, this  reduction can be translated into a relation between the scale factor ratio and the parameters $c_{\rm inf}$ and $c_{\rm UV}$ via
\be
\left(\frac{a_{c}}{a_{\rm end}}\right)^{n}=\left(\frac{c_{\rm inf}}{c_{\rm UV}}\right)^{\frac{1}{|\alpha|}}\,,
\label{eq:ratio1}
\ee
assuming $c_V=1$.
For a quadratic inflaton potential after the end of inflation, $V_{\rm inf}(\chi)\propto\chi^{2}$, the inflaton energy density scales as $V_{\rm inf}(\chi)\propto c_{\rm inf}^{\frac{8}{3}}\propto a^{-3}$. Therefore, during the interval $a_{\rm end}\lesssim a\lesssim a_{c}$, the ratio $c_{\rm inf}/c_{\rm UV}$ evolves according to
\be
\frac{c_{\rm inf}}{c_{\rm UV}}=\left(\frac{a_{c}}{a_{\rm end}}\right)^{\frac{9}{8}}\,.
\label{eq:ratio2}
\ee
Combining (\ref{eq:ratio1}) and (\ref{eq:ratio2}), we obtain the approximate estimate
\be
n\simeq\frac{9}{8|\alpha|}\,.
\label{eq:mexponent}
\ee
For the benchmark choice $|\alpha|=0.3$, this gives $n\simeq3.75$.

By substituting Eqs.~(\ref{eq:amaf}) and (\ref{eq:mexponent}) into Eq.~(\ref{eq:Pisohalf}), we obtain the following conservative estimate for the small-scale isocurvature amplitude 
in our scenario
\be
\mathcal{P}_{\rm iso}^{\frac{1}{2}}(a_{m},k_{f_a})\lesssim\frac{H_{\rm inf}}{2\pi f_{a}}\left(\frac{f_{\rm inf}}{f_{a}}\right)^{1-\frac{20}{9}|\alpha|}\frac{\Lambda_{\rm QCD}}{T_{\rm RH}}\,.
\label{eq:Pisohalf2}
\ee
For example, taking $\alpha=-0.3$, $H_{\rm inf}\simeq 5\times10^{13}\,\mathrm{GeV}$, $f_{a}=10^{9}\,\mathrm{GeV}$, and $f_{\rm inf}\simeq 10^{16}\,\mathrm{GeV}$, this gives
\be
\mathcal{P}_{\rm iso}^{\frac{1}{2}}(a_{m},k_{f_a})\lesssim 
\frac{3\times10^5\,{\rm GeV}}{T_{\rm RH}}\,.
\label{eq:Pisohalf3}
\ee
Hence, achieving sufficiently small isocurvature perturbations on small scales, such that $\mathcal{P}_{\rm iso}^{\frac{1}{2}}(a_{m},k_{f_a})\ll 2\pi$, requires
\be
T_{\rm RH}\gtrsim3\times10^5\,\mathrm{GeV}\,.
\label{eq:TRH6}
\ee
This requirement is easily compatible with the maximal reheating temperature,
\be
T_{\rm RH,max}\simeq\left(\frac{90}{\pi^{2}g_{*}}\right)^{1/4}\sqrt{H_{\rm inf}M_{4}}\simeq 10^{15}\,\mathrm{GeV}\,,
\ee
where the numerical value is obtained assuming the supersymmetric Standard Model with $g_{*}\simeq 230$. Therefore, provided that $T_{\rm RH}$ is not too low, our mechanism to suppress $\beta_{\rm iso}$ at CMB scales does not lead to domain-wall formation in the post-inflationary era.\footnote{In a supersymmetric scenario, an additional constraint on $T_{\rm RH}$ may arise if the extra-dimensional axion constitutes most of the dark matter. For high-scale supersymmetry breaking, $m_{3/2}>10^{3}\,{\rm TeV}$, gravitinos decay before BBN but still produce an LSP abundance $\Omega_{\rm LSP}h^{2}\approx0.1(T_{\rm RH}/10^{10}\,{\rm GeV})(m_{\rm LSP}/100\,{\rm GeV})$~\cite{Moroi:1993mb,Bolz:2000fu,Pradler:2006qh}. Requiring this contribution to remain below $\mathcal{O}(1)\%$ of the observed dark matter abundance for $m_{\rm LSP}=\mathcal{O}(100)\,{\rm GeV}$ gives $T_{\rm RH}\lesssim 10^{9}{\rm GeV}$.}

Finally, since the decay-constant transition induces a nonzero angular velocity, $\dot{\theta}_{a}\neq0$, one should also check that the associated axion kinetic energy, $\rho_{\theta}\simeq f_{a}^{2}\dot{\theta}_{a}^{2}$ does not temporarily dominate the total energy density of the Universe, $3M_{4}^{2}H^2(a_{f_a})$, near $a\sim a_{f_{a}}$. Using (\ref{eq:Pisohalf}) together with $\dot{\theta}_{a}\approx H(a_{f_a})\,\mathcal{P}_{\rm iso}^{1/2}(a_{f_{a}},k_{f_a})$, the condition $\rho_{\theta}\ll 3M_{4}^{2}H^2(a_{f_a})$ becomes
\be
f_{a}\,\mathcal{P}_{\rm iso}^{1/2}(a_{f_{a}},k_{f_a})\ll M_{4}\,.
\label{eq:energydominance}
\ee
For $f_{a}$ in the canonical QCD axion window, we have numerically verified that the condition (\ref{eq:energydominance}) is readily satisfied for $\alpha=-0.3$, $H_{\rm inf}\lesssim5\times10^{13}\,\mathrm{GeV}$, and $f_{\rm inf}\lesssim M_{4}$.

%%%%%%%%%%%%%%%%%%%%%%%%%%%%%%%%%

\section{Conclusion}
\label{sec:conclusion}

Extra-dimensional axions are well motivated by the axion quality problem, but if present during inflation they generically produce pre-inflationary isocurvature perturbations. We have proposed a dynamical suppression mechanism in warped extra dimensions, where the inflaton-radion dynamics reduce the inter-brane separation during inflation. The resulting larger warp factor enhances the axion decay constant during inflation relative to its late-time value, suppressing axion fluctuations while preserving radion stabilization and slow-roll inflation.

We derived the consistency conditions required for this scenario to be viable. The radion must remain sufficiently heavy compared to the Hubble scale during inflation, and the radion-induced corrections must not spoil slow roll. These conditions constrain
the dimensionless parameters controlling the warped geometry. In the benchmark region, the inflationary decay constant can reach values as large as $f_{\rm inf} = \mathcal{O}(10^{16})\,{\rm GeV}$ while the late-time decay constant remains in the canonical QCD axion window $10^{9}\,{\rm GeV}\lesssim f_a\lesssim10^{12}\,{\rm GeV}$.

We then examined the resulting implications for axion isocurvature perturbations. The enhancement of the decay constant during inflation suppresses the angular fluctuation of the axion field at horizon crossing, thereby reducing the isocurvature amplitude. For a mild hierarchy between $M_{4}$ and AdS curvature $k$, corresponding to $b_{\rm CFT}\lesssim 10$, our analysis shows that for inflationary scales $H_{\rm inf} \lesssim 10^{11}\,{\rm GeV}$ the resulting isocurvature fraction can be
compatible with the observational bound $\beta_{\rm iso}<0.036$ throughout much
of the QCD axion window, without requiring entropy dilution of the axion abundance or a fine-tuned initial misalignment angle. For $H_{\rm inf}\simeq 10^{12}\,{\rm GeV}$, values $f_{a}
\lesssim\mathcal{O}(10^{11}){\rm GeV}$ can be compatible with $\beta_{\rm iso}<0.036$. In this sense, the dynamical size of the extra dimension reopens a substantial region of parameter space that would otherwise be excluded by the isocurvature constraint.

A potential concern in models with time-dependent decay constants is the enhancement of small-scale axion fluctuations during the transition from $f_{\rm inf}$ to $f_a$. We analyzed this effect and found that, although such modes can indeed be amplified, they do not trigger domain-wall formation provided the reheating temperature is sufficiently high. In particular, for the benchmark $\alpha=-0.3$, this requires $T_{\rm RH}\gtrsim\mathcal{O}(10^{5})\,\mathrm{GeV}$. Furthermore, the extra-dimensional axion in our framework is not subject to the same parametric-resonance problem that can arise in field-theoretic large-decay-constant models~\cite{Kasuya:1996ns,Kasuya:1997td,Kawasaki:2013iha}. This is because our setup contains no coherently
oscillating Peccei--Quinn radial mode, and the axion decay constant evolves
monotonically rather than oscillating.

In summary, the mechanism developed here shows that the cosmological evolution of the extra-dimensional geometry can make extra-dimensional axions compatible with CMB isocurvature bounds at high inflationary scales that would otherwise be
excluded. The key point is that the effective axion decay constant depends on
both the size of the extra dimension and the higher-dimensional gauge coupling,
and can therefore be enhanced during inflation relative to its late-time value. While our analysis has focused on a specific realization, namely a time-dependent compactification scale driven by nontrivial radion dynamics during inflation, the mechanism is more general. A similar suppression of isocurvature
perturbations could arise in any setup where the compactification scale, or
moduli controlling the higher-dimensional gauge coupling, evolve
cosmologically. This opens a broader class of viable high-scale inflationary
histories for extra-dimensional axions and deserves further study.

\medskip\noindent\textit{Acknowledgments\,---\,}%
This work was supported by the Department of Energy under Grant No. DE-SC0011842 at the University of Minnesota.

%%%%%%%%%%%%%%%%%%%%%%%%%%%%%%%%%%%%%%%%%%%%%%%%%%%%%%%%%%%%%%%%%%%%%%%%%%%%%%%%%%%%%%%%%%%%%%%%%%%%

\bibliographystyle{JHEP}
\bibliography{arxiv1}

%%%%%%%%%%%%%%%%%%%%%%%%%%%%%%%%%%%%%%%%%%%%%%%%%%%%%%%%%%%%%%%%%%%%%%%%%%%%%%%%%%%%%%%%%%%%%%%%%%%%

\end{document}